\documentclass[12pt,preprint,url]{aastex}
\usepackage{amsbsy}
\usepackage{amsmath}
\newcommand{\be}{\begin{equation}}
\newcommand{\ee}{\end{equation}}

\def\ltsima{$\; \buildrel < \over \sim \;$}

\def\lsim{\lower.5ex\hbox{\ltsima}}
\def\gtsima{$\; \buildrel > \over \sim \;$}
\def\gsim{\lower.5ex\hbox{\gtsima}}

\shorttitle{Spin wandering near accretion torque balance}
\shortauthors{Melatos}

\begin{document}
\title{Spin wandering of an accreting neutron star near torque balance:
 implications for gravitational wave searches}

\author{A. Melatos\altaffilmark{1,2}}

\email{amelatos@unimelb.edu.au}

\altaffiltext{1}{School of Physics, University of Melbourne,
 Parkville, VIC 3010, Australia}

\altaffiltext{2}{Australian Research Council Centre of Excellence
 for Gravitational Wave Discovery (OzGrav), 
 Parkville, VIC 3010, Australia}

\begin{abstract}
\noindent 
Torque balance between gravitational radiation reaction and magnetocentrifugal accretion
is one reason why accreting neutron stars rotate slower than centrifugal break-up.
Random processes in the accretion disk,
such as Rayleigh-Taylor instabilities and flicker noise,
drive the spin frequency $f_\ast$ and gravitational wave frequency 
$f_{\rm gw}\propto f_\ast$ to wander stochastically around their torque balance values.
Here it is shown theoretically, 
in terms of an idealized Ornstein-Uhlenbeck model,
that the standard deviation of $f_{\rm gw}$ is given by
$\sigma_{f_{\rm gw}}(h_0) \propto f_{\rm gw} h_0 D \beta T^{1/2}$,
or equivalently
$\sigma_{f_{\rm gw}}(F_{\rm X}) \propto f_{\rm gw}^{1/2} F_{\rm X}^{1/2} D \beta T^{1/2}$,
where $h_0$ is the characteristic wave strain,
$F_{\rm X}$ is the X-ray flux,
$D$ is the source distance,
and $\beta$ and $T$ are the fractional amplitude and autocorrelation time-scale
of the stochastic torque.
One can use $\sigma_{f_{\rm gw}}(h_0)$ and $\sigma_{f_{\rm gw}}(F_{\rm X})$
to check the physical consistency of a detection candidate in a gravitational wave search
by comparing with $\sigma_{f_{\rm gw}}$ observed.
Spin wandering impacts search design,
because it sets the maximum time interval $\max(T_{\rm coh})$,
during which the signal may be treated as coherent.
The Ornstein-Uhlenbeck calculation predicts
$10^{-1} \lesssim \max(T_{\rm coh}) / (1\, {\rm day}) \lesssim 10^3$,
with $3\times 10^{-2} \lesssim \beta T^{1/2} / (1 \, {\rm s^{1/2}} ) \lesssim 4$,
for Rayleigh-Taylor instabilities and flicker noise.
The predicted $\max(T_{\rm coh})$ is shorter typically than the year-long observation runs of 
audio-band interferometers operating today,
emphasizing the utility of semi-coherent search algorithms,
which divide the data into multiple coherent segments of duration $\max(T_{\rm coh})$.
\end{abstract}


\section{Introduction
 \label{sec:wan1}}
Persistent, quasimonochromatic gravitational wave signals from rapidly rotating,
accreting neutron stars are key targets for terrestrial, audio-band, long-baseline
interferometers such as the Laser Interferometer Gravitational Wave Observatory (LIGO)
\citep{aas15},
Virgo \citep{ace15},
and the Kamioka Gravitational Wave Detector (KAGRA)
\citep{aku21}.
Targets of this kind, classified as continuous wave sources,
radiate in response to a time-varying mass quadrupole moment,
e.g.\ a thermoelastic or magnetic mountain
\citep{ush00,mel05},
or a time-varying current quadrupole moment,
e.g.\ from stellar oscillations such as r-modes
\citep{and98,owe98,nay06,car19,don25}.
If the gravitational radiation reaction torque \citep{tho80}
balances the magnetocentrifugal accretion torque \citep{gho79},
the characteristic gravitational wave strain scales as 
$h_0 \propto F_{\rm X}^{1/2} f_\ast^{-1/2}$,
independently of the source distance,
where $F_{\rm X}$ and $f_\ast$ denote the X-ray flux and spin frequency respectively
\citep{pap78b,wag84,bil98,and05,wat08,has15,ril23}.
That is, brighter X-ray sources are better targets,
all else being equal.
Torque balance is consistent with the observation,
that accreting neutron stars
rotate significantly slower than the centrifugal break-up limit for many plausible 
equations of state \citep{coo94,pri11},
with $f_\ast \lesssim 1 \, {\rm kHz}$ measured from X-ray pulsations
and thermonuclear burst oscillations
\citep{cha03}.

Continuous wave searches are vitiated by even modest amounts of stochastic spin wandering,
e.g.\ due to accretion torque fluctuations
\citep{bay93,dek93,bil97,pat09,san17,yan17,mel23,ser23}.
The sensitivity implications differ for coherent and semi-coherent searches 
\citep{wat08,ril23,car25}.
In a coherent search,
one must track the signal phase accurately to within half a cycle over
a total of
$\sim 3 \times 10^{10} (T_{\rm obs} / 1 \, {\rm yr}) (f_{\rm gw} / 1 \, {\rm kHz})$
cycles,
where $T_{\rm obs}$ is the total observation time, 
and $f_{\rm gw}$ is the signal frequency.
Stated another way, in the Fourier domain,
the standard deviation
$\sigma_{f_{\rm gw}}
 = \langle (f_{\rm gw} - \langle f_{\rm gw} \rangle)^2 \rangle^{1/2}$
of the stochastic signal frequency must satisfy 
$\sigma_{f_{\rm gw}} \lesssim (2 T_{\rm obs})^{-1} 
 = 2\times 10^{-8} (T_{\rm obs}/ 1 \, {\rm yr})^{-1} \, {\rm Hz}$
to track the signal.
One staple coherent algorithm is the ${\cal F}$-statistic,
a maximum likelihood matched filter
\citep{jar98}.
In semi-coherent searches,
which are the focus of this paper,
the total observation is divided into $N_T$ blocks, 
each of duration $T_{\rm coh} = N_T^{-1} T_{\rm obs}$
(i.e.\ the coherence time).
Individual blocks are analyzed coherently with a matched filter, 
and the per-block detection statistics are summed incoherently,
without demanding phase continuity between consecutive blocks.
Semi-coherent searches tolerate $N_T$ times more spin wandering,
with
$\sigma_{f_{\rm gw}} \lesssim (2 T_{\rm coh})^{-1} = N_T (2 T_{\rm obs})^{-1}$,
but they pay a sensitivity penalty:
the minimum detectable wave strain scales as $h_0\propto T_{\rm obs}^{-1/2}$
for coherent searches and
$h_0 \propto (T_{\rm coh} T_{\rm obs})^{-1/4} = N_T^{1/4} T_{\rm obs}^{-1/2}$
for semi-coherent searches.
Many semi-coherent algorithms are in use,
e.g.\ based on cross-correlation methods,
double Fourier transforms, or hidden Markov models
\citep{dhu08,goe11,suv16,suv17}.
Several semi-coherent searches using LIGO-Virgo-KAGRA data
have targeted accreting neutron stars,
e.g.\ the low-mass X-ray binaries Scorpius X$-$1
and XTE J1751$-$305
\citep{abb17b,abb17c,abb17d,abb22c,abb22d,var25}
and 20 accreting millisecond X-ray pulsars
\citep{mid20,abb22}.

An important challenge when doing a semi-coherent search for an
accreting neutron star is to set $T_{\rm coh}$ optimally
to match the true spin wandering time-scale,
with $T_{\rm coh} \approx (2\sigma_{f_{\rm gw}})^{-1}$;
see Figure 13 in \citet{mel21} for example.
If the source pulsates persistently in X-rays,
one estimates $\sigma_{f_{\rm gw}}$ empirically
by measuring $f_\ast$ and
assuming that the gravitational-wave-emitting quadrupole is locked
approximately to the X-ray-emitting crust.
Many systems, however, 
including high-priority targets with relatively high $F_X$ 
like Scorpius X$-$1, do not pulsate in X-rays at all
or do so intermittently.
One is then obliged to estimate $\sigma_{f_{\rm gw}}$ theoretically.
For example, the published searches for Scorpius X$-$1 
referenced in the previous paragraph assume
$10 \leq T_{\rm coh} / (1\, {\rm day}) \leq 70$
based on analytic calculations assuming a randomly kicked rotor
\citep{sam14}
or Monte Carlo simulations of torque fluctuations calibrated against the
Rossi X-ray Timing Explorer (RXTE) light curve of Scorpius X$-$1
\citep{muk18}.

In this paper, we generalize previous theoretical calculations
of $\sigma_{f_{\rm gw}}$ by incorporating mean reversion caused by
the gravitational radiation reaction and magnetocentrifugal accretion torques.
Specifically, we model the time-dependent spin of the neutron star, $f_\ast(t)$,
in terms of an Ornstein-Uhlenbeck process,
linearized about the torque balance equilibrium,
which is driven by a stochastic torque of arbitrary physical origin
with idealized white-noise statistics.
The Ornstein-Uhlenbeck model has been applied successfully to
RXTE timing data for 24 accretion-powered pulsars in the Small Magellanic Cloud
to measure their magnetic moments, radiative efficiencies, 
and magnetocentrifugal fastness parameters
\citep{mel23,ole24b,chr25,ole25}.
It has also been applied to study torque reversals and retrograde accretion
in 4U 1626$-$67 by analyzing timing data from the 
Compton Gamma-Ray Observatory and Fermi Gamma-Ray Space Telescope
\citep{ole26a}.
It builds on pioneering empirical work modeling spin wandering as a 
Wiener process (i.e.\ without mean reversion) by \citet{muk18}
and complements recent studies of the secular behavior of $f_\ast(t)$
involving torque balance, magnetic mountains, and crustal breakage 
\citep{pag25}.
The paper is organized as follows.
In Section \ref{sec:wan2},
we formulate the Ornstein-Uhlenbeck model and calculate $\sigma_{f_{\rm gw}}$
in terms of properties of the stochastic torque.
In Section \ref{sec:wan3},
we estimate $\sigma_{f_{\rm gw}}$ approximately
in two illustrative and representative astrophysical scenarios:
accretion disk flicker noise 
\citep{lyu97},
and Rayleigh-Taylor instabilities at the disk-magnetosphere boundary
\citep{rom02,rom03,rom15,bli16}.
In Section \ref{sec:wan4},
we calculate $T_{\rm coh}$ and discuss
the implications for semi-coherent gravitational wave searches targeting
accreting neutron stars.

\section{Ornstein-Uhlenbeck model of spin wandering
 \label{sec:wan2}}
We start by formulating and solving the linear Ornstein-Uhlenbeck model
for small-amplitude spin wandering around magnetorotational equilibrium, 
driven by an arbitrary stochastic torque with white-noise statistics.
In Section \ref{sec:wan2a}, 
we write down a rotational equation of motion
and justify the idealized form of the net torque astrophysically.
In Section \ref{sec:wan2b},
we linearize the equation of motion
and solve it for the stationary probability density function (PDF)
of the angular velocity fluctuations.
In Section \ref{sec:wan2c}, 
we convert the Ornstein-Uhlenbeck solution into a formula relating $\sigma_{f_{\rm gw}}$,
the standard deviation of the gravitational wave signal frequency,
and $h_0$, the characteristic gravitational wave strain.
The formula $\sigma_{f_{\rm gw}}(h_0)$ is useful in practice,
when a semi-coherent gravitational wave search yields an above-threshold candidate,
for which $h_0$ is measureed.
One can then test whether the implied value of $\sigma_{f_{\rm gw}}$ 
is consistent with the amount of spin wandering actually observed in the search.
In Section \ref{sec:wan2d}, 
we convert the Ornstein-Uhlenbeck solution into an equivalent formula 
relating $\sigma_{f_{\rm gw}}$ and the X-ray flux $F_{\rm X}$
(or equivalently the mass accretion rate $\dot{M}_{\rm a}$).
The formula $\sigma_{f_{\rm gw}}(F_{\rm X})$ is useful in practice,
before a semi-coherent search is performed,
because it predicts theoretically the optimal coherence time-scale $T_{\rm coh}$,
as explained in Section \ref{sec:wan4}.

\subsection{Equations of motion
 \label{sec:wan2a}}
Consider a neutron star modeled as a rigid, spherical rotor,
with mass $M$, radius $R$, and moment of inertia $I$.
We neglect triaxiality and precession as a first approximation.
Let $\Omega(t)=2\pi f_\ast(t)$ be the angular velocity of the star at time $t$.
Angular momentum conservation implies
\begin{equation}
 I \frac{d\Omega}{dt}
 =
 N_{\rm a}(\Omega) + N_{\rm gw}(\Omega) + \xi(t)~,
\label{eq:wan1}
\end{equation}
where $N_{\rm a}$ denotes the magnetocentrifugal torque,
which can be positive or negative;
$N_{\rm gw}$ denotes the gravitational radiation reaction torque,
which is negative;
and $\xi(t)$ denotes the stochastic driving torque,
which fluctuates randomly and takes either sign instantaneously with equal probability.

The exact form of $N_{\rm a}$ in (\ref{eq:wan1}),
especially how it depends on $\Omega$, is uncertain at the time of writing;
it depends on the complicated microphysics of Rayleigh-Taylor and
Kelvin-Helmholtz instabilities at the disk-magnetosphere boundary
among other factors
\citep{rom03,rom15}.
In this paper, we adopt the widely used phenomenological formula
\citep{gho79}
\begin{equation}
 N_{\rm a} = \dot{M}_{\rm a} (G M R_{\rm a})^{1/2} (1 - \omega_{\rm a})~,
\label{eq:wan2}
\end{equation}
where $\dot{M}_{\rm a}$ is the mass accretion rate,
the Alfv\'{e}n radius $R_{\rm a}$ marks approximately the disk-magnetosphere boundary,
$\omega_{\rm a} = (R_{\rm a}/ R_{\rm c})^{3/2}$ is the fastness parameter,
and $R_{\rm c} = (GM)^{1/3} \Omega^{-2/3}$ is the corotation radius,
where disk material in a Kepler orbit corotates with the star.
One can express $R_{\rm a}$ in terms of $M$, $\dot{M}_{\rm a}$,
and the star's magnetic dipole moment $\mu$ \citep{gho79,kul13},
but the expression is not required in what follows.
We emphasize that there are many uncertainties in (\ref{eq:wan2}).
We mention four examples below by way of illustration
and refer the reader to \citet{sti25} for a fuller discussion.
First, one may replace the factor $1-\omega_{\rm a}$ in (\ref{eq:wan2})
with some other dimensionless factor
$n(\omega_{\rm a}) \neq 1 - \omega_{\rm a}$ satisfying $| n(\omega_{\rm a}) | \sim 1$.
Several plausible alternatives are reviewed systematically by \citet{sti25};
some are polynomials in $\omega_{\rm a}$, while others are not.
Second, 
mass transfer may not be conservative;
the inflow assumed in (\ref{eq:wan2}) may be accompanied
simultaneously by an outflow, modifying $N_{\rm a}$
\citep{mat05,mat08}.
Third, the disk-magnetosphere boundary has nonzero thickness,
and one should distinguish between $R_{\rm a}$ and the truncation radius
$R_{\rm t} \sim R_{\rm a}$, where the inner edge of the disk terminates.
\footnote{
\citet{sti25} showed that one can write 
$N_{\rm a} = \dot{M}_{\rm a} (GMR_{\rm t})^{1/2} n(\omega_{\rm t})$,
with $\omega_{\rm t} = (R_{\rm t}/R_{\rm c})^{3/2}$,
for several plausible magnetocentrifugal models.
}
Fourth, 
$R_{\rm t} \propto \Omega^{-1/5}$ depends on $\Omega$,
when the pitch angle of the sheared magnetic field at the disk-magnetosphere boundary
is taken into account
\citep{wan87},
leading to important phenomena such as episodic accretion and disk trapping
\citep{dan10,dan12,dan17}.
The above four effects lie outside the scope of this paper,
but it will be interesting to incorporate them in the future,
once richer X-ray and gravitational wave data become available.

The physical origin of the fluctuating torque $\xi(t)$ is also uncertain.
Two possibilities, 
namely Rayleigh-Taylor instabilities at the disk-magnetosphere boundary
\citep{rom03,rom15,bli16}
and disk flicker noise \citep{lyu97}, 
are discussed in Sections \ref{sec:wan3c} and \ref{sec:wan3d} respectively,
but other possibilities exist too,
e.g.\ nonconservative mass transfer
\citep{mar19}
or a transient accretion disk
\citep{sah12}.
As a first approximation, we assume in this paper that $\xi(t)$ obeys
delta-correlated, white-noise, Langevin statistics,
with $\langle \xi(t) \rangle =0$ and
$\langle \xi(t) \xi(t') \rangle = \sigma^2 \delta(t-t')$,
where $\sigma$ (units: ${\rm g \, cm^2 \, s^{-3/2}}$)
is proportional to the amplitude of the fluctuating torque,
and $\langle \dots \rangle$ denotes an ensemble average over random
realizations of the noise process.
White noise in $\xi(t)$ produces colored noise in $\Omega(t)$
through the filtering action of $N_{\rm a}(\Omega)$
and $N_{\rm gw}(\Omega)$ in (\ref{eq:wan1}), 
consistent with X-ray timing observations
\citep{bay93,dek93,bil97,ser23};
the same happens in standard Brownian motion
\citep{gar94}.
The Langevin approximation has proved successful when applied to modeling 
RXTE observations of spin wandering in 24 accretion-powered pulsars in the Small Magellanic Cloud
\citep{ole24b,chr25},
yielding the first measurements of their fastness histories $\omega_{\rm a}(t)$, 
and distinguishing the stable, ordered unstable, and chaotic unstable accretion regimes
predicted theoretically
\citep{bli16,ole25}.
It is straightforward to generalize $\xi(t)$ to colored noise,
if future X-ray timing observations warrant.

\subsection{Linearizing around torque balance
 \label{sec:wan2b}}
The angular velocity fluctuations of accreting neutron stars are small
in percentage terms
\citep{bil97}.
For example,
\citet{yan17} classified 14 out of 24 accretion-powered pulsars
in the Small Magellanic Cloud as being near rotational equilibrium,
in the arbitrary sense that their spin period derivatives 
$|dP/dt|$ are less than $1.5$ times the measurement error 
over $\sim 15 \, {\rm yr}$ of monitoring.
Letting $\sigma_P$ denote the standard deviation of $P=2\pi/\Omega$
around a least-squares fit with $dP/dt = {\rm constant}$,
one measures
$0.011 \leq \sigma_P / (1 \, {\rm s}) \leq 9.6$
and 
$1.4 \times 10^{-3} \leq \sigma_P / P \leq 1.1\times 10^{-2}$
for the 14 objects,
implying fluctuations $\lesssim 1\%$ of $\Omega$.
\footnote{
Measurements of $P$, $dP/dt$, and $\sigma_P$ for the 14 objects are quoted
in the first, second, and third columns respectively of Table 3 in \citet{yan17}.
}
Consequently, it is appropriate to treat the driver $\xi(t)$ in (\ref{eq:wan1}) 
as a first-order quantity and linearize (\ref{eq:wan1}) about the equilibrium
angular velocity $\Omega_{\rm eq}$.
Writing $\Omega(t) = \Omega_{\rm eq} + \Omega^{(1)}(t)$,
with 
$| \Omega^{(1)}(t) | \ll \Omega_{\rm eq}$, 
we obtain
\begin{equation}
 I \frac{d\Omega^{(1)}}{dt}
 =
 \left[
  \left( \frac{dN_{\rm a}}{d\Omega} \right)_{\rm eq} + 
  \left( \frac{dN_{\rm gw}}{d\Omega} \right)_{\rm eq} 
 \right]
 \Omega^{(1)}
 + \xi(t)~. 
\label{eq:wan3}
\end{equation}
The subscript `eq' in (\ref{eq:wan3}) means that the derivative is evaluated
at $\Omega=\Omega_{\rm eq}$.
The equilibrium angular velocity satisfies
$0 = N_{\rm a}(\Omega_{\rm eq}) + N_{\rm gw}(\Omega_{\rm eq})$,
i.e.\ torque balance 
\citep{pap78b,wag84,bil98}.
It can be expressed as
$\Omega_{\rm eq} = (GM)^{1/2} R_{\rm a}^{-3/2} \tilde{\Omega}$,
where $0< \tilde{\Omega} < 1$ (dimensionless)
denotes the unique positive real root of the quintic equation
\begin{equation}
 0 = 
 \frac{\dot{M}_{\rm a} R_{\rm a}^8 (1 - \tilde{\Omega}) }{K_{\rm gw} (GM)^2} 
 -
 \tilde{\Omega}^5~,
\label{eq:wan4}
\end{equation}
and we write $N_{\rm gw} = - K_{\rm gw} \Omega^5$
with $K_{\rm gw} \propto \epsilon^2 I^2$,
where $\epsilon$ denotes the mass ellipticity
\citep{jar98}.
\footnote{
Equation (\ref{eq:wan4}) assumes mass quadrupole gravitational radiation.
For current quadrupole radiation, one writes instead
$N_{\rm gw} = - K_{\rm gw,curr} \Omega^7$
and replaces (\ref{eq:wan4}) with
$0=\dot{M}_{\rm a} R_{\rm a}^{11} K_{\rm gw,curr}^{-1} (GM)^{-3} (1 - \tilde{\Omega})
-\tilde{\Omega}^7$.
\label{foot:wan3}
}

The stochastic differential equation (\ref{eq:wan3}) describes
an Ornstein-Uhlenbeck process, i.e.\ standard, mean-reverting, Brownian motion
\citep{gar94}.
Mean reversion is guaranteed,
because one has $(dN_{\rm a}/d\Omega)_{\rm eq} + (dN_{\rm gw}/d\Omega)_{\rm eq} < 0$
for all $\Omega_{\rm eq}$.
This makes sense physically: 
if the star spins up momentarily to $\Omega > \Omega_{\rm eq}$,
then $N_{\rm gw} \propto \Omega^5$ becomes more negative,
$\omega_{\rm a} \propto \Omega$ increases,
and $N_{\rm a}$ becomes less positive,
implying $d\Omega/dt <0$;
and the opposite happens,
if the star spins down momentarily to $\Omega < \Omega_{\rm eq}$.
\footnote{
Mean reversion occurs even for $N_{\rm gw}=0$:
the magnetocentrifugal torque acts alone to restore equilibrium
in both the accretion phase 
($\Omega < \Omega_{\rm eq}$, $\omega_{\rm a} < 1$, $N_{\rm a} > 0$)
and the propeller phase 
($\Omega > \Omega_{\rm eq}$, $\omega_{\rm a} > 1$, $N_{\rm a} < 0$)
\citep{gho79}.
The propeller phase is less relevant for continuous gravitational wave searches, 
whose targets satisfy $N_{\rm gw} \neq 0$ and are visible as accreting X-ray sources.
However, the situation is nuanced:
some matter falls onto the star even in the weak propeller regime 
$1 < \omega_{\rm a} \lesssim 1.25$
\citep{ust06,dan10,lii14,pap15,mel23,ole25}.
}
The solution of the Ornstein-Uhlenbeck process is well known
\citep{gar94}.
In the astrophysically relevant regime
$t \gg I | (dN_{\rm a}/d\Omega)_{\rm eq} + (dN_{\rm gw}/d\Omega)_{\rm eq} |^{-1}$,
after the initial transient dies away,
the process becomes stationary,
and the PDF of the angular velocity fluctuations approaches
\begin{equation}
 p[\Omega^{(1)} ]
 = 
 \left( 
  \frac{\gamma I}{\pi \sigma^2}
 \right)^{1/2}
 \exp\left\{
  - \frac{\gamma I [\Omega^{(1)}]^2}{\sigma^2}
 \right\}~.
\label{eq:wan5}
\end{equation}
In (\ref{eq:wan5}) and henceforth, we adopt the shorthand
\begin{equation}
 \gamma = | (dN_{\rm a}/d\Omega)_{\rm eq} + (dN_{\rm gw}/d\Omega)_{\rm eq} |
\label{eq:wan6}
\end{equation}
for the mean reversion rate.
Equation (\ref{eq:wan5}) implies
\begin{equation}
 \langle [\Omega^{(1)}]^2 \rangle = \sigma^2/(2\gamma I)~.
\label{eq:wan7}
\end{equation}
The small-amplitude approximation therefore holds for
$\sigma \ll \Omega_{\rm eq} (2 \gamma I)^{1/2}$.
Equation (\ref{eq:wan7}) implies $\sigma_{f_{\rm gw}} \propto \sigma$,
as shown in Appendix \ref{sec:wanappc} and Sections \ref{sec:wan2c} and \ref{sec:wan2d}.

\subsection{$\sigma_{f_{\rm gw}}$ as a function of $h_0$
 \label{sec:wan2c}}
When a semi-coherent continuous wave search discovers an above-threshold detection candidate, 
it returns measurements of the characteristic wave strain, $h_0$,
and the maximum drift in the signal frequency during the observation, $\Delta f_{\rm gw}$.
For example, algorithms based on a hidden Markov model assume typically,
that $f_{\rm gw}$ is piecewise-constant in each coherent block of duration $T_{\rm coh}$,
and reconstruct the optimal frequency path $f_{\rm gw}(t)$ from one block to the next
by dynamic programming, whereupon one calculates
$\Delta f_{\rm gw} = \max_{0\leq t,t'\leq T_{\rm obs}} |f_{\rm gw} (t)-f_{\rm gw}(t') |$
\citep{suv16,suv17}.
The stack-slide algorithm operates in a similar way
\citep{bra00,pri12}.
\footnote{
Semi-coherent algorithms that assume $f_{\rm gw}={\rm constant}$ 
for $0\leq t \leq T_{\rm obs}$
and exclude spin wandering a priori are not discussed in this paper;
see \citet{car25} for a detailed comparative study.
}
With $h_0$ and $\Delta f_{\rm gw}$ in hand,
one can check the detection candidate for physical consistency.
If the target is an isolated neutron star, 
$h_0$ implies a minimum, secular decrement in $f_{\rm gw}$ due to $N_{\rm gw}$.
If the predicted decrement exceeds the observed $\Delta f_{\rm gw}$,
the candidate is unlikely to be astrophysical.
If the target is an accreting neutron star, on the other hand,
$N_{\rm gw}$ and $N_{\rm a}$ may be in torque balance,
and the foregoing argument does not apply.
Instead, we apply the results in Section \ref{sec:wan2b}
to formulate an alternative consistency check:
is the observed $\Delta f_{\rm gw}$ consistent with $\langle [\Omega^{(1)}]^2 \rangle$,
as calculated from (\ref{eq:wan5})?
Note that $\langle [\Omega^{(1)}]^2 \rangle$ decreases at fixed $\sigma$,
as $h_0$ and hence $\gamma$ increase.

Continuous wave searches favor high-$F_{\rm X}$ targets,
such as Scorpius X$-$1, because torque balance predicts $h_0 \propto F_{\rm X}^{1/2}$
\citep{pap78b,wag84,bil98}.
In Appendix \ref{sec:wanappa},
we show that the mean reversion rate $\gamma$ in (\ref{eq:wan6}) is dominated by 
$|(dN_{\rm gw}/d\Omega)_{\rm eq}|$,
unless the system is fine-tuned to satisfy $5/6 \leq \omega_{\rm a} < 1$,
which is unlikely astrophysically for a high-$F_{\rm X}$ target;
see Appendix \ref{sec:wanappa} for a detailed justification.
Hence, in the absence of fine-tuning,
we obtain 
$\gamma \approx 
|(dN_{\rm gw}/d\Omega)_{\rm eq}|=5\Omega_{\rm eq}^{-1} |N_{\rm gw}|$.
Moreover, the radiation reaction torque acting on 
an orthogonal, biaxial rotor, which does not precess,
can be related (e.g.\ via the radiated power) 
to $h_0$ and the source distance $D$ through
\begin{equation}
 | N_{\rm gw} |
 =
 \frac{2 c^3 D^2 h_0^2 \Omega}{5G}~;
\label{eq:wan8}
\end{equation}
see equations (14) and (23) in \citet{zim80} and \citet{jar98} respectively,
as well as equations (14) and (16) in \citet{ril23}.
\footnote{
The counterpart of (\ref{eq:wan8}) for current quadrupole radiation from r-modes
follows from (21) and (23) in \citet{ril23} and takes the same form
as (\ref{eq:wan8}),
except that the factor $2/5$ is replaced by $8/45$
for a canonical r-mode with $\Omega=3\pi f_{\rm gw}/2$,
i.e.\ $f_{\rm gw} = 4f_\ast/3$ without relativistic and equation of state
corrections 
\citep{and98,owe98,car19}.
}
Upon combining (\ref{eq:wan7}), (\ref{eq:wan8}),
$\gamma = 5\Omega_{\rm eq}^{-1} |N_{\rm gw}|$,
and
$\Omega= \pi f_{\rm gw}$ for an orthogonal, biaxial rotor,
we arrive at
\begin{equation}
 \sigma_{f_{\rm gw}}^2
 =
 \frac{G \sigma^2}{4\pi^2 c^3 I D^2 h_0^2}~.
\label{eq:wan9}
\end{equation}

Equation (\ref{eq:wan9}) fulfills the goal articulated at the start of this section.
When a semi-coherent continuous wave search discovers an above-threshold detection candidate,
accompanied by measurements of $\Delta f_{\rm gw}$ and $h_0$,
one can apply (\ref{eq:wan9}) to test for physical consistency.
If $\Delta f_{\rm gw}$ and $\sigma_{f_{\rm gw}}(h_0)$ differ greatly,
the candidate is unlikely to be astrophysical.
Beyond $\Delta f_{\rm gw}$ and $h_0$,
the test requires some knowledge (albeit approximate) of $\sigma$, $I$, and $D$.
The source distance $D$ is measured accurately for some but not all objects,
e.g.\ to $\approx 10\%$ accuracy via trigonometric parallax at radio wavelengths 
for Scorpius X$-$1
\citep{bra99}.
The moment of inertia depends on the equation of state,
with $I \sim 10^{45} \, {\rm g \, cm^2}$ fiducially
\citep{lat07}.
What remains is to evaluate $\sigma$ for specific, plausible mechanisms 
relevant to accreting neutron stars targeted for continuous wave searches.
This is done for Rayleigh-Taylor instabilities and disk flicker noise
in Sections \ref{sec:wan3c} and \ref{sec:wan3d} respectively.

\subsection{$\sigma_{f_{\rm gw}}$ as a function of $F_{\rm X}$
 \label{sec:wan2d}}
Before a continuous wave search begins,
it is important to estimate $\sigma_{f_{\rm gw}}$ on astrophysical grounds
in order to select $T_{\rm coh}$, as discussed in Section \ref{sec:wan1}.
Equation (\ref{eq:wan9}) is inadequate for the task:
one does not know $h_0$, before the search begins.
If torque balance applies, however,
one can rewrite (\ref{eq:wan9}) in terms of the known X-ray flux $F_{\rm X}$
instead of $h_0$.
The resulting formula, $\sigma_{f_{\rm gw}}(F_{\rm X})$,
can be used to ensure that a signal is tracked coherently
within each block of duration $T_{\rm coh}$,
by requiring that the frequency bin width, 
$(2 T_{\rm coh})^{-1}$,
exceeds $\sigma_{f_{\rm gw}}(F_{\rm X})$.

The calculation proceeds as in Section \ref{sec:wan2c}.
For high-$F_{\rm X}$ targets accreting near the Eddington rate,
the arguments in Appendix \ref{sec:wanappa} imply
$\omega_{\rm a} \ll 1$ and hence
$\gamma \approx 
|(dN_{\rm gw}/d\Omega)_{\rm eq}|=5\Omega_{\rm eq}^{-1} |N_{\rm gw}|
 = 5\Omega_{\rm eq}^{-1} |N_{\rm a}|$,
where the latter equality follows from torque balance.
Upon substituting the expression for $\gamma$ into (\ref{eq:wan7})
and eliminating $\dot{M}_{\rm a}$ using (\ref{eq:wan2}) and
\begin{equation}
 F_{\rm X} = \frac{G M \dot{M}_{\rm a}}{4\pi D^2 R}~,
\label{eq:wan10}
\end{equation}
we obtain
\begin{equation}
 \sigma_{f_{\rm gw}}^2
 = 
 \frac{(GM)^{1/2} \Omega_{\rm eq} \sigma^2}{40\pi^3 I D^2 R R_{\rm a}^{1/2} F_{\rm X}}~.
\label{eq:wan11}
\end{equation}
Equation (\ref{eq:wan11}) is evaluated conservatively by letting
$\sigma_{f_{\rm gw}}$ be as large as reasonably possible.
This is achieved by taking $R_{\rm a} \approx R$
(consistent with $\omega_{\rm a} \ll 1$)
and setting $\Omega_{\rm eq} = \pi f_{\rm gw}$ to match the top of the band
($f_{\rm gw} \lesssim 0.5 \, {\rm kHz}$).
\footnote{
In the special case of a detection candidate,
for which $F_{\rm X}$ and $h_0$ are both measured accurately,
one can solve for $R_{\rm a}$ from (\ref{eq:wan4}) independently of $D$ in principle;
see Appendix \ref{sec:wanappb}.
}
The assumed equation of state determines $M$, $R$, and $I$
\citep{lat07}.
As with (\ref{eq:wan9}), what then remains is to estimate $\sigma$
for plausible stochastic mechanisms; see Section \ref{sec:wan3}.

\section{Stochastic torque
 \label{sec:wan3}}
In general, it is hard to predict the noise amplitude $\sigma$ from first principles.
For example, no predictive theory exists to describe timing noise in an isolated radio pulsar,
e.g.\ due to stochastic processes in its superfluid interior
\citep{cor85,jon90,mel14b}.
In accreting systems, however, spin wandering is thought to be driven
predominantly by fluctuations in the hydromagnetic accretion torque,
making it possible to relate $\sigma$ to 
$N_{\rm a}(\Omega_{\rm eq})$ and hence $N_{\rm gw}(\Omega_{\rm eq})=-N_{\rm a}(\Omega_{\rm eq})$
(torque balance).
In Section \ref{sec:wan3a},
we relate $\sigma$ to the autocorrelation time-scale $T$ of $N_{\rm a}$.
In Section \ref{sec:wan3b},
we reexpress $\sigma_{f_{\rm gw}}$ in (\ref{eq:wan9}) and (\ref{eq:wan11}) 
in terms of $T$ instead of $\sigma$.
We then evaluate $T$ for two specific stochastic processes in the literature:
Rayleigh-Taylor instabilities at the disk-magnetosphere boundary
(Section \ref{sec:wan3c}),
and disk flicker noise
(Section \ref{sec:wan3d}).

\subsection{Autocorrelation time-scale
 \label{sec:wan3a}}
Let us consider first the Ornstein-Uhlenbeck process described by (\ref{eq:wan3})
over a short time interval $t_0 \leq t' \leq t_0+t$, with $t \ll \gamma^{-1}$,
so that both $\Omega^{(1)}(t')$ and the Langevin driver $\xi(t')$ behave as Wiener
processes to a good approximation.
The random angular momentum increment during the interval is given by
$I \Omega^{(1)}(t+t_0) - I \Omega^{(1)}(t_0) 
 = \int_{t_0}^{t+t_0} dt' \, \xi(t')$,
and its variance equals $\sigma^2 t$
\citep{gar94}.
Equivalently, consider instead an idealized microphysical model,
in which $\xi(t)$ remains constant, with value $\pm \beta N_{\rm a}(\Omega_{\rm eq})$,
during successive time intervals of duration $T$
and switches sign randomly from one interval to the next.
In the microphysical picture, 
there are $t/T$ intervals of duration $T$ during $t_0 \leq t' \leq t_0+t$, 
and the variance of the angular momentum increment equals
$(t/T)[\beta N_{\rm a}(\Omega_{\rm eq}) T]^2$.
Equating the two equivalent variances, we arrive at the standard result
\citep{gar94}
\begin{equation}
 \sigma^2
 =
 \beta^2 N_{\rm a}(\Omega_{\rm eq})^2 T~.
\label{eq:wan12}
\end{equation}
In (\ref{eq:wan12}), $T$ is the autocorrelation time-scale of $\xi(t)$
and satisfies $T \ll \gamma^{-1}$ in the Langevin (white noise) regime,
while $\beta$ is a dimensionless constant of order unity to be quantified below
(see Sections \ref{sec:wan3c} and \ref{sec:wan3d}),
which characterizes what fraction of $N_{\rm a}(\Omega_{\rm eq})$ fluctuates
in the specific hydromagnetic mechanism under consideration.

\subsection{$\sigma_{f_{\rm gw}}$ as a function of $h_0$ or $F_{\rm X}$
 \label{sec:wan3b}}
We use (\ref{eq:wan12}) to eliminate $\sigma$ in favor of $\beta$ and $T$
in (\ref{eq:wan9}) and (\ref{eq:wan11}) assuming torque balance,
i.e.\ $| N_{\rm a}(\Omega_{\rm eq}) | = | N_{\rm gw}(\Omega_{\rm eq}) |$.
Armed with a detection candidate discovered by a continuous wave search,
for which $h_0$ and $f_{\rm gw}$ are measured,
we use (\ref{eq:wan9}) to calculate
\begin{eqnarray}
 \sigma_{f_{\rm gw}}^2
 & = &
 \frac{c^3 \beta^2 T f_{\rm gw}^2 h_0^2 D^2}{25 G I}
\label{eq:wan13}
 \\
 & = &
 1.5 \times 10^{-11}
 \left( \frac{f_{\rm gw}}{0.1 \, {\rm kHz}} \right)^2
 \left( \frac{h_0}{10^{-26}} \right)^2
 \left( \frac{D}{10 \, {\rm kpc}} \right)^2
\nonumber
 \\
 & & \times
 \left( \frac{I}{10^{45} \, {\rm g \, cm^2}} \right)^{-1}
 \left( \frac{\beta}{1} \right)^2
 \left( \frac{T}{1 \, {\rm s}} \right)
 \, {\rm Hz^2}~.
\label{eq:wan14}
\end{eqnarray}
Armed with an X-ray flux measurement, before a continuous wave search begins,
we use (\ref{eq:wan11}) to calculate
\begin{eqnarray}
 \sigma_{f_{\rm gw}}^2
 & = &
 \frac{2 \beta^2 T f_{\rm gw} R R_{\rm a}^{1/2} F_{\rm X} D^2}{5 I (GM)^{1/2}}
\label{eq:wan15}
 \\
 & = &
 2.8 \times 10^{-11}
 \left( \frac{f_{\rm gw}}{0.1 \, {\rm kHz}} \right)
 \left( \frac{R}{10^6 \, {\rm cm}} \right)
 \left( \frac{R_{\rm a}}{10^6 \, {\rm cm}} \right)^{1/2}
 \left( \frac{F_{\rm X}}{10^{-8} \, {\rm erg \, cm^{-2} \, s^{-1}}} \right)
 \left( \frac{D}{10 \, {\rm kpc}} \right)^2
\nonumber
 \\
 & & \times
 \left( \frac{I}{10^{45} \, {\rm g \, cm^2}} \right)^{-1}
 \left( \frac{M}{1.4 M_\odot} \right)^{-1/2}
 \left( \frac{\beta}{1} \right)^2
 \left( \frac{T}{1 \, {\rm s}} \right)
 \, {\rm Hz^2}~.
\label{eq:wan16}
\end{eqnarray}

The fiducial estimates (\ref{eq:wan14}) and (\ref{eq:wan16}),
viz.\ $\sigma_{f_{\rm gw}} \lesssim 10^{-5} \, {\rm Hz}$,
match or exceed the typical frequency bin width
$(2 T_{\rm coh})^{-1}=6\times 10^{-7} (T_{\rm coh}/ 10\, {\rm days})^{-1} \, {\rm Hz}$
of semi-coherent searches published previously
\citep{abb22d,abb22,abb22c},
confirming expectations that spin wandering is appreciable under
plausible conditions
\citep{wat08,ril23}.
The fiducial values of $\beta$ and $T$ in (\ref{eq:wan14}) and (\ref{eq:wan16}) 
are motivated in Sections \ref{sec:wan3c} and \ref{sec:wan3d}
for Rayleigh-Taylor instabilities and disk flicker noise respectively.

\subsection{Rayleigh-Taylor instabilities
 \label{sec:wan3c}}
Global, three-dimensional, magnetohydrodynamic simulations of disk accretion 
onto a rapidly rotating, magnetized, compact object reveal,
that the boundary between the accretion disk and magnetosphere is unstable
\citep{rom02,rom03,kul08,kul13,bli16};
see \citet{lai14} and \citet{rom15} for comprehensive reviews.
The unstable dynamics involve a complicated mixture of Kelvin-Helmholtz modes,
driven by the shear between the Keplerian disk and corotating magnetosphere for $\omega_{\rm a} \neq 1$;
and Rayleigh-Taylor modes,
which occur because the magnetosphere acts as a lighter fluid
supporting the heavier disk under gravity
\citep{aro76}.
Nonlinear feedback loops regulate the instabilities,
so that their dynamics are mean-reverting and approximately cyclic,
i.e.\ self-healing.
\footnote{
See Section 2.4 in \citet{mel23} for a detailed discussion of self-healing instabilities
in the context of an idealized, Ornstein-Uhlenbeck model of magnetocentrifugal accretion
in X-ray binaries.
}
An instability grows at one or more locations on the disk-magnetosphere boundary,
one or more transient magnetic channels open up which connect the disk to the star,
``fingers'' of disk material break through the magnetosphere and strike the stellar surface,
then accretion replenishes the mass reservoir at the boundary,
the magnetic channels shut,
and equilibrium is restored temporarily
\citep{dan10,dan12,dan17}.
With respect to Rayleigh-Taylor instabilities in particular,
\citet{bli16} identified three distinct accretion regimes:
stable (two stationary magnetospheric funnel flows and hotspots), 
chaotic unstable (multiple migrating funnels and hotspots), 
and ordered unstable (two migrating funnels and hotspots),
corresponding to $\omega_{\rm a} \gtrsim 0.6$,
$0.45 \lesssim \omega_{\rm a} \lesssim 0.6$,
and
$\omega_{\rm a} \lesssim 0.45$ respectively.
A Kalman filter analysis of RXTE pulse timing data for 24 accretion-powered pulsars
in the Small Magellanic Cloud confirms observationally the existence of the regimes
\citep{ole25}.
Motivated thus, we focus on Rayleigh-Taylor instabilities in this section,
while noting that other complicated mechanisms operate simultaneously,
e.g.\ an outflowing wind
\citep{mat05,mat08}.

Simulations like those referenced above demonstrate,
that the accretion torque fluctuates by a factor of order unity
about its mean value, i.e.\ one has $\beta \sim 1$.
This is visible clearly in Figure 9 in \citet{rom02}
and Figure 21 in \citet{rom03}, for example.
Moreover, the time-step in the simulations is short enough to resolve
the fluctuations.
One finds that they are autocorrelated on a characteristic time-scale
$\approx 5\pi^{-1}$ times the Kepler period at the inner edge of the disk,
\footnote{
For example, \citet{rom02} defined the simulation time unit to be 
$t_0=R_0 (GM/R_0)^{-1/2}$ and set $R_0 = R_{\rm t}$ initially;
see Section 2.1 of the latter reference.
The fluctuations are autocorrelated visually on the time-scale $T\approx 10 t_0$;
see Figure 9 of the latter reference.
}
viz.\ $T\approx 5\pi^{-1}T_{\rm K}(R_{\rm t}) \approx 5\pi^{-1} T_{\rm K}(R_{\rm a})$,
implying 
\begin{equation}
 T = 7 \times 10^{-4}
 \left( \frac{M}{1.4 M_\odot} \right)^{-1/2}
 \left( \frac{R_{\rm a}}{10^6 \, {\rm cm}} \right)^{3/2}
 \, {\rm s}~.
\label{eq:wan17}
\end{equation}
Upon substituting (\ref{eq:wan17}) into (\ref{eq:wan14}) or (\ref{eq:wan16}),
one finds $\sigma_{f_{\rm gw}} \lesssim 10^{-7} \, {\rm Hz}$.
The result is interesting.
It suggests that spin wandering driven by Rayleigh-Taylor instabilities 
alone may be too weak to affect semi-coherent searches,
with $T_{\rm coh} \lesssim 10 \, {\rm days}$,
but it may affect coherent searches,
with $T_{\rm coh} = T_{\rm obs} \sim 1 \, {\rm yr}$.

We emphasize that there are many uncertainties in (17).
The physics of the disk-magnetosphere boundary is complicated by factors
which cannot yet be predicted analytically from first principles,
e.g.\ the specific, twisted geometry of the magnetic field,
together with factors that are challenging to address even with state-of-the-art
numerical simulations,
e.g.\ the extreme dynamic range of the turbulent processes involved,
such as magnetic reconnection
\citep{lai14,rom15}.
Disk warping and precession also occur,
when the magnetic and rotation axes are misaligned
\citep{lai99,fou11,rom21}.
High-resolution X-ray timing observations provide valuable constraints,
but they are hard to interpret uniquely without imaging the 
disk-magnetosphere boundary spatially,
which is not feasible for the foreseeable future.
For all these reasons and more,
equation (\ref{eq:wan17}) should be treated as an order-of-magnitude estimate.
One should factor in a conservative safety margin when using (\ref{eq:wan17})
to make design decisions about continuous wave searches,
e.g.\ when selecting $T_{\rm coh}$ (see Section \ref{sec:wan4}).

\subsection{Disk flicker noise
 \label{sec:wan3d}}
The accretion rate $\dot{M}_{\rm a}$ and hence $N_{\rm a}$ fluctuate
on time-scales longer than $T_{\rm K}(R_{\rm a})$,
driven by turbulent hydromagnetic processes in the outer disk 
away from the disk-magnetosphere boundary.
One important example is disk flicker noise
\citep{lyu97}.
It is driven by spatiotemporal fluctuations of the $\alpha$ parameter 
describing the effective disk viscosity
\citep{sha73}.
Disk flicker noise and Rayleigh-Taylor instabilities are uncorrelated
in general so they add to the noise amplitude $\sigma$ in quadrature.

Global, three-dimensional, magnetohydrodynamic simulations like those
referenced in Section \ref{sec:wan3c} do not run long enough to track
the disk flicker noise time-scale.
Instead, we rely on analytic predictions
\citep{lyu97},
that $T$ is comparable to the viscous or accretion time-scale
at the radius $R_{\rm f} \gg R_{\rm a}$,
where the $\alpha$ fluctuations originate,
i.e.
\begin{eqnarray}
 T & = &
 (2\pi \alpha)^{-1} (\tan\theta_{\rm d})^{-2} T_{\rm K}(R_{\rm f})
\label{eq:wan18}
 \\
 & = &
 2\times 10^{1} 
 \left( \frac{\alpha}{10^{-2}} \right)^{-1}
 \left( \frac{\tan\theta_{\rm d}}{10^{-1}} \right)^{-2}
 \left( \frac{M}{1.4 M_\odot} \right)^{-1/2}
 \left( \frac{R_{\rm a}}{10^6 \, {\rm cm}} \right)^{3/2}
 \left( \frac{R_{\rm f}}{10R_{\rm a}} \right)^{3/2}
 \, {\rm s}~;
\label{eq:wan19}
\end{eqnarray}
see equation (2) in \citet{lyu97}.
In (\ref{eq:wan18}),
$\alpha\ll 1$ denotes the standard Shakura-Sunyaev disk viscosity
\citep{sha73},
and $\theta_{\rm d}$ is the disk opening angle.
The fractional amplitude $\beta$ is challenging to predict
for disk flicker noise but satisfies $\beta \sim 1$
\citep{lyu97}.
The spectrum of $\dot{M}_{\rm a}$ fluctuations is expected to follow
a power law, whose properties are determined partly by the balance between
viscous heating and radiative cooling;
see Sections 2--4 in \citet{lyu97}.
The foregoing order-of-magnitude estimates of $\beta$ and $T$ 
are consistent with empirical studies of $F_{\rm X}$
fluctuations in X-ray pulsars,
which imply $\beta \sim 0.2$ and $T \lesssim 10^7 {\rm s}$
\citep{muk18,ser22};
see also Section 5 in \citet{ole24b}.

Upon comparing (\ref{eq:wan17}) and (\ref{eq:wan18}),
one sees that $T$ and hence $\sigma^2$ are greater 
for disk flicker noise than for Rayleigh-Taylor instabilities.
For a thin disk 
\footnote{
Equation (\ref{eq:wan18}) can be modified to describe a thick,
advection-dominated disk,
by setting $\tan\theta_{\rm d}=1$;
see (22) in \citet{lyu97}.
}
with $\theta_{\rm d} \lesssim 0.1$ 
and $R_{\rm f} \gtrsim 10R_{\rm a}$,
one obtains $T \sim 10^{-3} \, {\rm s}$ and $T \gtrsim 10^1 \, {\rm s}$
for the two mechanisms respectively.
Equations (\ref{eq:wan14}) and (\ref{eq:wan16}) then imply
$\sigma_{f_{\rm gw}} \gtrsim 10^{-5} \, {\rm Hz}$
for disk flicker noise, 
i.e.\ $\sim 10^2$ times greater than for Rayleigh-Taylor instabilities.
Hence spin wandering driven by disk flicker noise is likely to affect
semi-coherent searches, with $T_{\rm coh} \lesssim 10 \, {\rm days}$,
as well as coherent searches, with $T_{\rm coh} = T_{\rm obs} \sim 1\, {\rm yr}$.
We discuss this issue further in the following section.

\section{Gravitational wave searches
 \label{sec:wan4}}
A key decision when designing a semi-coherent continuous wave search is
how to choose $N_T$, the number of coherent blocks,
and hence $T_{\rm coh}=N_T^{-1} T_{\rm obs}$, the duration of each coherent block.
The sensitivity scales as 
$h_0 \propto (T_{\rm coh} T_{\rm obs})^{-1/4} = N_T^{1/4} T_{\rm obs}^{-1/2}$
\citep{dhu08,wat08},
so it makes sense to keep $N_T$ as low as possible, ceteris paribus.
However, one must ensure that the spin wandering amplitude $\sigma_{f_{\rm gw}}$
does not exceed the frequency bin width in each coherent block,
so that the signal stays within one bin as it wanders
and hence is tracked coherently over the interval $T_{\rm coh}$.
This translates into the condition $\sigma_{f_{\rm gw}} \leq (2 T_{\rm coh})^{-1}$,
or alternatively
\begin{equation}
 \max (T_{\rm coh})
 =
 5.8 
 \left( \frac{\sigma_{f_{\rm gw}}}{10^{-6} \, {\rm Hz}} \right)^{-1}
 \, {\rm days}~.
\label{eq:wan20}
\end{equation}
In (\ref{eq:wan20}), $\sigma_{f_{\rm gw}}$ is given by (\ref{eq:wan16})
together with (\ref{eq:wan17}) or (\ref{eq:wan19}) depending on the physical origin
of the stochastic torque.
Equation (\ref{eq:wan20}) is broadly consistent with previous estimates
based on analytic calculations assuming a randomly kicked rotor
without mean reversion 
\citep{sam14},
Monte Carlo simulations calibrated against the power spectral density 
of $F_{\rm X}$ fluctuations measured by RXTE for Scorpius X$-$1
\citep{muk18},
and studies of torque balance accompanied by crustal breakage
\citep{pag25}.

Ideally, one measures $\sigma_{f_{\rm gw}}$ directly from the electromagnetic ephemeris,
if the source pulsates.
\footnote{
Even when the source pulsates, there is no guarantee that the phase of the
gravitational-wave-emitting quadrupole is tied exactly to the rotational phase
of the stellar crust and hence the pulse phase,
so typically one scans a narrow band around the electromagnetic ephemeris,
e.g.\ $|f_{\rm gw} - 2 f_\ast | \leq 2\times 10^{-3} f_\ast$
\citep{aba25}.
}
Unfortunately, many accreting neutron stars prioritized as search targets
by the LIGO-Virgo-KAGRA collaboration,
e.g.\ low-mass X-ray binaries with relatively high $F_{\rm X}$,
do not pulsate
\citep{wat08,ril23,wet23}.
Under such circumstances, one is obliged to estimate $\sigma_{f_{\rm gw}}$
from first principles.
Equation (\ref{eq:wan16}) in this paper offers one way to do so,
using measurements of $F_{\rm X}$ and $D$,
combined with estimates of $T$ from (\ref{eq:wan17}) and (\ref{eq:wan19})
for Rayleigh-Taylor instabilities and disk flicker noise respectively.
As discussed in Sections \ref{sec:wan3c} and \ref{sec:wan3d},
the formula (\ref{eq:wan20}) implies $\max(T_{\rm coh}) \sim 10^2\, {\rm days}$
and $\max(T_{\rm coh}) \sim 1 \, {\rm day}$ for the two mechanisms
respectively, noting of course that the estimates contain uncertainties
and should be applied with a safety margin.

\begin{figure}[ht]
\begin{center}
\includegraphics[width=15cm,angle=0]{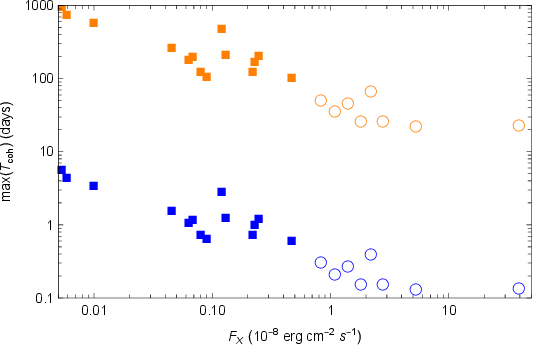}
\end{center}
\caption{
Maximum duration of a coherent block, $\max(T_{\rm coh})$ (units: days),
as a function of time-averaged X-ray flux, 
$F_{\rm X}$ (units: ${\rm erg \, cm^{-2} \, s^{-1}}$),
for a representative sample of 22 accreting neutron stars tabulated by \citet{wat08},
calculated from (\ref{eq:wan16}), (\ref{eq:wan17}), (\ref{eq:wan19}), and (\ref{eq:wan20})
to ensure that a semi-coherent continuous wave search 
tracks spin wandering successfully.
The orange and blue symbols correspond to spin wandering driven by
Rayleigh-Taylor instabilities and disk flicker noise respectively.
The filled squares and open circles correspond to sources featuring
thermonuclear burst oscillations and kilohertz quasiperiodic oscillations
respectively.
Measurements of $F_{\rm X}$ and $D$ are drawn from Table 1 in \citet{wat08}.
Fiducial parameters:
$f_{\rm gw} = 0.1 \, {\rm kHz}$,
$R=R_{\rm a} = 0.1 R_{\rm f} = 10^6 \, {\rm cm}$,
$I = 10^{45} \, {\rm g \, cm^2}$,
$M = 1.4 M_\odot$, $\beta=1$, $\alpha=10^{-2}$, $\tan\theta_{\rm d} = 0.1$.
}
\label{fig:wan1}
\end{figure}

Figure \ref{fig:wan1} illustrates the practical implications of (\ref{eq:wan20})
for semi-coherent continuous wave searches.
It graphs $\max( T_{\rm coh})$, predicted from (\ref{eq:wan20}),
as a function of the measured, long-term, time-averaged X-ray flux $F_{\rm X}$,
for 22 representative accreting neutron stars studied as promising search targets
by \citet{wat08};
see also \citet{ril23}.
The targets divide into sources that exhibit thermonuclear burst oscillations
(filled squares; 14 sources)
and kilohertz quasiperiodic oscillations
(open circles; eight sources).
\footnote{
Table 1 in \citet{wat08} lists three outliers in the two categories
with $F_{\rm X} \leq 5 \times 10^{-11} \, {\rm erg \, cm^{-2} \, s^{-1}}$
and $\max(T_{\rm coh}) \geq 10^3 \, {\rm days}$.
They are not plotted in Figure \ref{fig:wan1},
because they are unlikely to be detectable for the foreseeable future
assuming torque balance and $T_{\rm obs} \sim 1 \, {\rm yr}$.
}
The orange and blue symbols correspond to
Rayleigh-Taylor instabilities and disk flicker noise respectively;
see (\ref{eq:wan17}) and (\ref{eq:wan19}).
The main trend is straightforward to interpret:
$\max(T_{\rm coh}) \propto \sigma_{f_{\rm gw}}^{-1} \propto (F_{\rm X} D^2)^{-1/2}$ decreases,
as $F_{\rm X}$ increases, 
because the $N_{\rm a}$ fluctuations increase,
and shorter coherent blocks are needed to track the signal. 
The trend is not monotonic, because different targets have different $D$.
Interestingly, if Rayleigh-Taylor instabilities are the only driver 
of spin wandering,
one finds $10 \lesssim \max(T_{\rm coh}) / (1 \, {\rm day}) \lesssim 10^2$
and hence $3 \lesssim N_T \lesssim 30$ for quasiperiodic oscillation sources,
assuming $T_{\rm obs} = 1\, {\rm yr}$ for the sake of definiteness.
Burst oscillation sources are tracked with $N_T \lesssim 3$
and even $N_T=1$ (fully coherent) in some instances,
noting that it is prudent to maintain a safety margin.
On the other hand, if disk flicker noise drives spin wandering too,
$N_T$ rises substantially,
and fully coherent searches are ruled out for the objects in the figure.
Specifically, 
one finds $0.1 \lesssim \max(T_{\rm coh}) / (1 \, {\rm day}) \lesssim 10$
and hence $3\times 10^1 \lesssim N_T \lesssim 3\times 10^3$,
accompanied by a sensitivity penalty $\propto N_T^{-1/4}$
relative to a fully coherent search (see Section \ref{sec:wan1}).
For some semi-coherent algorithms,
e.g.\ based on hidden Markov models \citep{suv16,suv17}, 
$T_{\rm coh}$ is bounded below by twice the length of the
short Fourier transforms which constitute the input data format
\citep{ast05,wet23}.
The latter condition yields $T_{\rm coh} \geq 1 \, {\rm hr}$ typically,
which approaches the lower edge of Figure \ref{fig:wan1}.

None of the objects in Figure \ref{fig:wan1} pulsate persistently, 
so it is impossible to measure $\sigma_{f_{\rm gw}}$ directly and independently 
at the time of writing.
An accreting millisecond X-ray pulsar does not suffer from this drawback.
Its electromagnetic ephemeris is measured accurately,
and an upper bound on $\sigma_{f_{\rm gw}}$ can be measured directly,
without resorting to theoretical estimates such as (\ref{eq:wan16});
see, for example, the published searches using a hidden Markov model 
to analyze five objects in the second LIGO observing run
\citep{mid20}
and 20 objects in the third LIGO observing run
\citep{abb22}.
The upper bound on $\sigma_{f_{\rm gw}}$ measured directly from X-ray timing,
viz.\ $\sigma_{f_{\rm gw}} \lesssim 10^{-7} \, {\rm Hz}$
\citep{wat08},
is broadly consistent with (\ref{eq:wan16}) and (\ref{eq:wan17})
for Rayleigh-Taylor instabilities
with $F_{\rm X} \leq 2 \times 10^{-10} \, {\rm erg \, cm^{-2} \, s^{-1}}$
\citep{wat08}.
\footnote{
Aquila X$-$1 is an exception,
with $F_{\rm X} \approx 1.2 \times 10^{-9} \, {\rm erg \, cm^{-2} \, s^{-1}}$
and $\sigma_{f_{\rm gw}} \approx 9\times 10^{-4} \, {\rm Hz}$
\citep{wat08}.
}
Disk flicker noise predicts higher $\sigma_{f_{\rm gw}}$ values,
but it is important to note that the disk-magnetosphere physics
and the accretion regime are not necessarily the same
in accreting millisecond X-ray pulsars and the objects plotted in Figure \ref{fig:wan1},
some of which may not be in torque balance.
We also note that $f_\ast$ is not measured reliably
for several objects in Figure \ref{fig:wan1};
some burst oscillation frequencies are detected tentatively or in just one outburst,
and the difference frequency of the twin quasiperiodic oscillations
may not relate cleanly to $f_\ast$
\citep{wat08}.
Therefore, for the sake of definiteness when constructing Figure \ref{fig:wan1},
we put all 22 objects at the same fiducial $f_{\rm gw} = 0.1 \, {\rm kHz}$.
The reader is encouraged to experiment with alternatives.

\section{Conclusion
 \label{sec:wan5}}
Accreting neutron stars are high-priority targets for
continuous wave searches with audio-band, long-baseline, gravitational wave interferometers
\citep{ril23,wet23}.
Under the assumption that the gravitational radiation reaction torque
balances the magnetocentrifugal accretion torque,
higher priority is accorded to objects with higher X-ray flux $F_{\rm X}$,
because the characteristic gravitational wave strain scales as 
$h_0 \propto F_{\rm X}^{1/2} f_\ast^{-1/2}$
\citep{pap78b,wag84,bil98}.
However, random processes in the accretion disk cause the 
signal frequency $f_{\rm gw} \propto f_\ast$ to wander stochastically,
especially when $F_{\rm X}$ is relatively high
\citep{wat08}.
Spin wandering limits the maximum duration 
$\max(T_{\rm coh})$ of a coherent block in a semi-coherent continuous wave search,
because the standard deviation of the wandering signal frequency,
$\sigma_{f_{\rm gw}}$,
must stay below the frequency bin width $(2T_{\rm coh})^{-1}$
to track the signal successfully.

In this paper, we calculate $\sigma_{f_{\rm gw}}$ and hence $\max(T_{\rm coh})$
from first principles in the context of an idealized Ornstein-Uhlenbeck model
of spin wandering.
The model generalizes previous pioneering analyses
\citep{sam14,muk18,pag25}
by including mean reversion,
as the system fluctuates around torque balance.
It is shown (see Appendix \ref{sec:wanappa})
that the radiation reaction torque dominates the mean reversion rate $\gamma$ typically.
Two practical analytic formulas for $\sigma_{f_{\rm gw}}$ are derived
in terms of the fractional amplitude $\beta$ and autocorrelation time-scale $T$
of the torque fluctuations.
(i) Equation (\ref{eq:wan14}) predicts 
$\sigma_{f_{\rm gw}}(h_0)^2 \propto f_{\rm gw}^2 h_0^2 D^2 \beta^2 T$.
It is useful, when a semi-coherent search returns one or more detection candidates,
each with measured values of $f_{\rm gw}$, $h_0$, $D$, and $\Delta f_{\rm gw}$.
One screens the candidate for physical consistency by checking
that it satisfies $\Delta f_{\rm gw} \approx \sigma_{f_{\rm gw}}(h_0)$;
see Section \ref{sec:wan2c}.
(ii) Equation (\ref{eq:wan16}) predicts
$\sigma_{f_{\rm gw}}(F_{\rm X})^2 \propto f_{\rm gw} F_{\rm X} D^2 \beta^2 T$.
It is useful, before a semi-coherent search begins,
because it allows the data analyst to estimate $\max(T_{\rm coh})$ based on
electromagnetic measurements of $F_{\rm X}$, $D$, and possibly $f_{\rm gw}$
(or bounds on $f_{\rm gw}$ set by the detector's sensitivity);
see Section \ref{sec:wan2d}.
It is also useful for screening detection candidates for physical consistency
by checking
$\Delta f_{\rm gw} \approx \sigma_{f_{\rm gw}}(F_{\rm X})$,
once $\Delta f_{\rm gw}$ is measured gravitationally.
The noise parameters $\beta$ and $T$ are challenging to predict 
from first principles in general.
They are estimated for Rayleigh-Taylor instabilities at the disk-magnetosphere boundary
and disk flicker noise in (\ref{eq:wan17}) and (\ref{eq:wan19}) respectively,
drawing on magnetohydrodynamic simulations
\citep{rom02,rom03,rom15}
and analytic theory
\citep{lyu97}
established in the literature.

Figure \ref{fig:wan1} illustrates quantitatively the implications for gravitational wave searches
with reference to 22 promising and representative targets studied by \citet{wat08}.
If Rayleigh-Taylor instabilities drive spin wandering predominantly
(orange symbols),
targets that exhibit quasiperiodic oscillations are tracked reliably
by a semi-coherent search with 
$10 \lesssim \max(T_{\rm coh}) / (1 \, {\rm day}) \lesssim 10^2$
and hence $3 \lesssim N_T \lesssim 30$,
assuming $T_{\rm obs} = 1\, {\rm yr}$ for the sake of definiteness.
Targets that exhibit burst oscillations require $N_T \lesssim 3$
(nearly fully coherent).
On the other hand, if disk flicker noise drives spin wandering too
(blue symbols),
one finds $0.1 \lesssim \max(T_{\rm coh}) / (1 \, {\rm day}) \lesssim 10$
and hence $3\times 10^1 \lesssim N_T \lesssim 3\times 10^3$
for both oscillation categories.
The rise in $N_T$ is accompanied by a sensitivity penalty $\propto N_T^{-1/4}$
relative to a fully coherent search (see Section \ref{sec:wan1}).
Whatever scenario the data analyst prefers,
it is prudent to maintain a safety margin.

There are many ways that this paper may be refined in the future.
First and foremost, the Ornstein-Uhlenbeck model is idealized.
The phenomenological formula (\ref{eq:wan2}) for the magnetocentrifugal torque
omits important physics related to outflows \citep{mat05,mat08},
magnetic shear in the disk-magnetosphere boundary \citep{wan87}
and its role in disk trapping \citep{dan10,dan12,dan17},
and disk warping and precession \citep{lai99,fou11,rom21},
which are discussed in Section \ref{sec:wan2a}.
Many alternative forms of (\ref{eq:wan2}) are viable
\citep{sti25}.
Second, the $\beta$-$T$ parameterization of the stochastic torque
in Section \ref{sec:wan3a} is idealized,
e.g.\ the ensemble statistics of $\xi(t)$ are likely to be colored
rather than white in reality.
Good prospects exist,
to infer a better model of the stochastic torque 
(including better estimates of $\beta$ and $T$)
by analyzing high-resolution time series of $F_{\rm X}$ and $P$
with a Kalman filter,
as has been demonstrated successfully with RXTE data 
from 24 accretion-powered pulsars in the Small Magellanic Cloud
\citep{yan17,mel23,ole24b,chr25,ole25}.
A systematic study in this direction lies outside the scope of this paper
and will require additional data for some of the targets referenced by \citet{wat08}.
Finally, the specific accretion history of each target matters.
For example, many accreting neutron stars experience X-ray outbursts;
see the two rightmost columns of Table 1 in \citet{wat08}.
Outbursts can run for weeks in burst oscillation sources
and boost $F_{\rm X} \propto \dot{M}_{\rm a}$ temporarily by $\gtrsim 10$ times,
thereby increasing $\sigma_{f_{\rm gw}}$ and reducing $\max (T_{\rm coh})$.

\acknowledgments
AM thanks Nicholas Low and Andres Vargas for stimulating discussions.
This work is supported by the Australian Research Council 
Centre of Excellence for Gravitational Wave Discovery
(grant number CE230100016).

\appendix
\section{Physical connection between $\sigma_{f_{\rm gw}}$ and $\sigma$
 \label{sec:wanappc}}
The key expressions for $\sigma_{f_{\rm gw}}$ in terms of $h_0$ and $F_{\rm X}$
in Sections \ref{sec:wan2c} and \ref{sec:wan2d},
viz.\ equations (\ref{eq:wan9}) and (\ref{eq:wan11}) respectively,
exhibit the important proportionality $\sigma_{f_{\rm gw}} \propto \sigma$.
In this appendix, we explain how the proportionality arises physically.

The autocorrelation function of the white noise torque $\xi(t)$ defined
in Section \ref{sec:wan2a} satisfies 
$\langle \xi(t) \xi(t') \rangle = \sigma^2 \delta(t-t')$,
where $\langle \dots \rangle$ denotes an ensemble average
over random realizations of the noise
\citep{gar94}.
In the absence of mean reversion (i.e.\ $\gamma=0$),
the angular velocity fluctuation driven by $\xi(t)$ is given by
\begin{equation}
 \Omega^{(1)}(t)
 =
 I^{-1} \int_0^t dt' \, \xi(t')~,
\label{eq:wanappc1}
\end{equation}
with 
$\Omega^{(1)}(t) = \Omega(t) - \langle \Omega(t) \rangle = \Omega(t) - \Omega_{\rm eq}$.
Hence the variance of $\Omega^{(1)}(t)$ equals
\begin{equation}
 \langle [ \Omega^{(1)}(t) ]^2 \rangle 
 =
 \sigma^2 t / I^2~.
\label{eq:wanappc2}
\end{equation}
That is, for $\gamma=0$, the root mean square amplitude of the angular velocity fluctuation
grows in proportion to $\sigma$ and increases monotonically $\propto t^{1/2}$.
In the presence of mean reversion (i.e.\ $\gamma\neq 0$),
the angular velocity fluctuation is given instead by
\begin{equation}
 \Omega^{(1)}(t)
 =
 I^{-1} e^{-\gamma t/I} \int_0^t dt' \, e^{\gamma t' / I} \xi(t')~,
\label{eq:wanappc3}
\end{equation}
and the variance of $\Omega^{(1)}(t)$ saturates for $t \gg I / (2\gamma)$
at the stationary value
\begin{equation}
 \langle [ \Omega^{(1)}(t) ]^2 \rangle 
 =
 \sigma^2 / (2\gamma I)~,
\label{eq:wanappc4}
\end{equation}
in line with (\ref{eq:wan7}).
That is, for $\gamma \neq 0$, the root mean square amplitude of the angular velocity fluctuation
again grows in proportion to $\sigma$.

In this paper, we assume that the star rotates rigidly.
We neglect for simplicity any differential rotation between the solid crust, the superfluid core,
and the gravitational-wave-emitting quadrupole
\citep{bay69,has15,mey21a,don26}.
Hence the frequency of the gravitational wave signal is directly proportional to the
spin frequency of the star, with
$f_{\rm gw} = \eta \Omega/(2\pi)$,
$\langle f_{\rm gw} \rangle = \eta \Omega_{\rm eq}/(2\pi)$,
and hence
\begin{eqnarray}
 \sigma_{f_{\rm gw}}
 & = &
 \langle (f_{\rm gw} - \langle f_{\rm gw} \rangle)^2 \rangle^{1/2}
\label{eq:wanappc5}
 \\
 & = &
 \frac{\eta\sigma}{2 \pi (2\gamma I)^{1/2}}~
\label{eq:wanappc6}
\end{eqnarray}
from (\ref{eq:wanappc4}).
The dimensionless proportionality constant $\eta$ depends on the multipole
of the gravitational radiation.
For mass quadrupole radiation, e.g.\ from a mountain,
one has $\eta=2$ for an orthogonal, biaxial rotor
\citep{jar98},
$\eta=1$ and $\eta=2$ for an oblique, biaxial rotor
\citep{jar98},
and $\eta \approx 1$ and $\eta\approx 2$ 
(with corrections of order the mass ellipticity $\epsilon$)
for a triaxial rotor
\citep{zim80,van05,las13}.
For current quadrupole radiation, e.g.\ from r-modes,
one has $\eta \approx 4/3$
\citep{and98,owe98}
with appreciable corrections arising from general relativistic effects
and the equation of state
\citep{loc00,ste01,idr15,car19}.
In this paper, we adopt $\eta = 2$ for the sake of definiteness.
It is straightforward to generalize the results to other $\eta$ values if desired.

\section{Mean reversion rate
 \label{sec:wanappa}}
In this appendix, we show that the mean reversion rate
$\gamma = | (dN_{\rm a}/d\Omega)_{\rm eq} + (dN_{\rm gw}/d\Omega)_{\rm eq} |$
in (\ref{eq:wan6}) is dominated by 
$| (dN_{\rm gw}/d\Omega)_{\rm eq} |$,
unless the system is fine-tuned to satisfy $5/6 \leq \omega_{\rm a} <1$.
Note that $(dN_{\rm a}/d\Omega)_{\rm eq}$ and $(dN_{\rm gw}/d\Omega)_{\rm eq}$
are both negative, as torque balance requires $0 < \tilde{\Omega} < 1$.
They contribute additively to $\gamma$,
yielding 
$\gamma \geq \max [ |(dN_{\rm a}/d\Omega)_{\rm eq}| , |(dN_{\rm gw}/d\Omega)_{\rm eq}| ]$.

With respect to the magnetocentrifugal contribution to $\gamma$, 
equation (\ref{sec:wan2}) implies
\begin{equation}
 \left(
  \frac{dN_{\rm a}}{d\Omega}
 \right)_{\rm eq}
 =
 -\dot{M}_{\rm a} R_{\rm a}^2~,
\label{eq:wanappa1}
\end{equation}
which is independent of $\Omega_{\rm eq}$.
In reality, 
$(dN_{\rm a}/d\Omega)_{\rm eq}$ inherits a weak dependence on $\Omega_{\rm eq}$
from the disk truncation radius $R_{\rm t} \propto \Omega_{\rm eq}^{-1/5}$,
with $R_{\rm t} \sim R_{\rm a}$
\citep{wan87,dan10,dan12,dan17,sti25},
but we neglect it as a first approximation in this paper;
see Section \ref{sec:wan2a}.
With respect to the gravitational radiation reaction contribution,
$N_{\rm gw} \propto \Omega_{\rm eq}^5$ implies
\begin{equation}
 \left(
  \frac{dN_{\rm gw}}{d\Omega}
 \right)_{\rm eq}
 =
 \frac{5N_{\rm gw}(\Omega_{\rm eq})}{\Omega_{\rm eq}}~,
\label{eq:wanappa2}
\end{equation}
which is proportional to $\Omega_{\rm eq}^4$.
It is easy to convert (\ref{eq:wanappa2}) from mass to current quadrupole radiation
by replacing the factor 5 with 7;
see footnote \ref{foot:wan3}.
Upon setting $N_{\rm gw}(\Omega_{\rm eq}) = -N_{\rm a}(\Omega_{\rm eq})$
in (\ref{eq:wanappa2}) to reflect torque balance
and writing $\Omega_{\rm eq}=R_{\rm c}^{-3/2} (GM)^{1/2}$,
we obtain
\begin{equation}
 \frac{ |(dN_{\rm a}/d\Omega)_{\rm eq}| }{ |(dN_{\rm gw}/d\Omega)_{\rm eq}| }
 =
 \frac{\omega_{\rm a}}{5(1-\omega_{\rm a})}~.
\label{eq:wanappa3}
\end{equation}

Equation (\ref{eq:wanappa3}) demonstrates that the radiation reaction
contribution to $\gamma$ exceeds the magnetocentrifugal contribution,
unless the system satisfies the fine-tuning condition
$5/6\leq \omega_{\rm a} < 1$.
Observations reveal that a minority of accreting neutron stars occupy 
this fastness interval persistently.
For example, a Kalman filter analysis of RXTE data 
(viz.\ $P$ and $F_{\rm X}$ time series)
from 24 accretion-powered pulsars in the Small Magellanic Cloud
found that 
(i) 14 of the objects satisfy $\omega_{\rm a}(t) \lesssim 0.45$ 
for $\geq 98\%$ of the observation interval $0\leq t \leq T_{\rm obs}$;
and (ii) the other 10 objects roam peripatetically through the range
$0.6 \lesssim \omega_{\rm a}(t) \lesssim 1$
for $\geq 70\%$ of $0\leq t \leq T_{\rm obs}$
\citep{ole25}.
Moreover, when we specialize to high-$F_{\rm X}$ targets 
for continuous gravitational wave searches,
the regime $\omega_{\rm a}(t) \lesssim 0.45$ is favored strongly
for two reasons.
First, high-$F_{\rm X}$ sources accrete near the Eddington rate
($\dot{M}_{\rm a} \approx 1\times 10^{-8} M_\odot \, {\rm yr^{-1}}$)
and hence satisfy $\omega_{\rm a} \ll 1$,
if the stellar magnetic moment satisfies
$\mu \ll 4\times 10^{27} (M/1.4 M_\odot)^{5/6} (f_\ast/ 0.1 \, {\rm kHz})^{-7/6}
 (\dot{M}_{\rm a} / 10^{-8} M_\odot \, {\rm yr^{-1}})^{1/2}
 \, {\rm G\, cm^3}$
\citep{gho79,sti25},
which holds approximately for low-mass X-ray binaries
\citep{taa86,shi89,pay04,zha06}.
Second, 
searches with audio-band long-baseline interferometers
are restricted currently to $0.05 \lesssim f_{\rm gw} / (1 \, {\rm kHz}) \lesssim 0.5$
by Newtonian and quantum noise at the lower and upper ends of the band respectively
\citep{aas15,ace15,aku21}.
This guarantees $\omega_{\rm a} \leq 0.45$
and hence
$|(dN_{\rm a}/d\Omega)_{\rm eq}| / |(dN_{\rm gw}/d\Omega)_{\rm eq}| \leq 0.16$
for
$\mu \leq 5\times 10^{26} (M/1.4 M_\odot)^{5/6}
 (\dot{M}_{\rm a} / 10^{-8} M_\odot \, {\rm yr^{-1}})^{1/2}
 \, {\rm G\, cm^3}$
and $f_{\rm gw} = 2 f_\ast$
\citep{jar98}.

\section{Multimessenger measurement of the Alfv\'{e}n radius
 \label{sec:wanappb}}
When a semi-coherent gravitational wave search yields a detection candidate,
it returns measurements of the characteristic gravitational wave strain $h_0$
and signal frequency $f_{\rm gw}$.
In this appendix, we show how to combine $h_0$ and $f_{\rm gw}$ 
with a measurement of the X-ray flux $F_{\rm X}$
and the physical assumption of torque balance
to measure the Alfv\'{e}n radius $R_{\rm a}$.

Consider the torque balance condition in the form (\ref{eq:wan4}).
Let us eliminate $\dot{M}_{\rm a}$ in favor of $F_{\rm X}$
through (\ref{eq:wan10})
and $K_{\rm gw} = \Omega_{\rm eq}^{-5} |N_{\rm gw}|$
in favor of $h_0$ through (\ref{eq:wan8}).
Equation (\ref{eq:wan4}) then reduces to
\begin{equation}
 0 =
 \frac{10\pi G R F_{\rm X} \tilde{\Omega}^{1/3}(1-\tilde{\Omega})}
  {c^3 (GM)^{1/3} \Omega_{\rm eq}^{4/3} h_0^2}
 - 1~.
\label{eq:wanappb1}
\end{equation}
Equation (\ref{eq:wanappb1}) is a quartic in the dimensionless variable
$\tilde{\Omega}^{1/3} = (GM)^{-1/6} R_{\rm a}^{1/2} \Omega_{\rm eq}^{1/3}$.
Hence, if $\Omega_{\rm eq} = \pi f_{\rm gw}$ is measured,
equation (\ref{eq:wanappb1}) can be solved for $R_{\rm a}$.
Indeed (\ref{eq:wanappb1}) is a quartic in $R_{\rm a}^{1/2}$
and takes the explicit form
\begin{equation}
 0 =
 4\pi R F_{\rm X} R_{\rm a}^{1/2} 
  [ 1 - \pi f_{\rm gw} (GM)^{-1/2} R_{\rm a}^{3/2} ]
 -
 \frac{2\pi c^3 (GM)^{1/2} f_{\rm gw} h_0^2}{5 G}~.
\label{eq:wanappb2}
\end{equation}

Equation (\ref{eq:wanappb2}) is independent of the source distance $D$.
It does depend on $M$, $R$, and hence the equation of state
\citep{lat07},
consistent with (\ref{eq:wan9}) and (\ref{eq:wan11}).
It will be interesting to use future gravitational wave detections
to test how the multimessenger measurement of $R_{\rm a}$
from (\ref{eq:wanappb1}) or (\ref{eq:wanappb2}) compares with
predictions from phenomenological theories of magnetocentrifugal accretion
\citep{gho79,sti25},
magnetohydrodynamic simulations
\citep{kul13},
and independent measurements based on X-ray pulse timing data,
e.g.\ when analyzed with a Kalman filter
\citep{mel23,ole24b,chr25,ole25}.

\bibliographystyle{mn2e}
\bibliography{globular}

\end{document}